\documentclass[12pt]{article}
\usepackage{amssymb,amsmath,epsfig}

\begin{document}
\title{\bf Analysis of $f(R)$ Theory Corresponding to NADE and NHDE}
\author{M. Sharif \thanks{msharif.math@pu.edu.pk} and M. Zubair
\thanks{mzubairkk@gmail.com}\\
Department of Mathematics, University of the Punjab,\\
Quaid-e-Azam Campus, Lahore-54590, Pakistan.}

\date{}

\maketitle

\begin{abstract}
We develop the connection of $f(R)$ theory with new agegraphic and
holographic dark energy models. The function $f(R)$ is reconstructed,
regarding the $f(R)$ theory as an effective description for these dark energy
models. We show the future evolution of $f$ and conclude that these functions
represent distinct pictures of cosmological eras. The cosmological parameters
such as equation of state parameter, deceleration parameter, statefinder
diagnostic and $\omega-\omega'$ analysis are investigated which assure the
evolutionary paradigm of $f$.
\end{abstract}

\section{Introduction}

Expanding paradigm of the universe has been affirmed by the
contemporary observational data \cite{1}. The prime source behind
this dramatic change in the evolution of the universe is said to be
\emph{dark energy} (DE). Dark energy is a strange type of
gravitationally repulsive energy component, spread over 72\% of the
contents in the universe. The nature of DE is still a question mark
and various representations have been proposed in general theory of
relativity to understand it. Holographic dark energy (HDE) appeared
as one of the most eminent candidates to address the issue of cosmic
acceleration. The density of the HDE has been proposed by
incorporating the mathematical form of the holographic principle as
\cite{2,3}
\begin{equation*}
\rho_{\vartheta}=3c^2M^{2}_pL^{-2},
\end{equation*}
where $c$ is a constant, $M_p^{-2}=8\pi{G}$ is the reduced Planck
mass and $L$ is the infrared (IR) cutoff. Though Hubble horizon
$H^{-1}$ is the natural possibility for $L$ but it does not imply
the cosmic acceleration \cite{4}. Li \cite{3} suggested that the
future event horizon is the most appropriate choice for IR cutoff
which seems to be consistent with recent measurements.

The modification of the IR cutoff in HDE has been reported in
different scenarios such as introducing new time scale, considering
$L$ as a function of the Ricci scalar in both original and
generalized form. Wei and Cai \cite{5} suggested a new model of
agegraphic DE by introducing conformal time as the time scale for
the FRW universe and is known as new agegraphic DE (NADE). Wu et al.
\cite{6} discussed the evolution of the new agegraphic quintessence
DE models in $\omega-\omega'$ phase plane both with and without
interaction. The NADE model has been formulated in the context of
alternative theories such as Brans-Dicke theory \cite{7} and
Ho$\check{r}$ava-Lifshitz gravity \cite{8}. Granda and Oliveros
\cite{9} proposed a new IR cutoff for HDE in terms of $H$ and
$\dot{H}$ and discussed the correspondence of new HDE (NHDE) with
models of scalar fields. This work has been extended for interacting
case in non-flat universe \cite{10}.

The modification of the Einstein-Hilbert action is another promising
approach to explain the fact of cosmic acceleration. In this regard,
there are various theories of gravity such as $f(R)$ \cite{11},
$f(R,T)$, where $T$ is the trace of the energy-momentum tensor
\cite{12}-\cite{16a}, Gauss-Bonnet gravity \cite{17} etc. The action
of $f(R)$ theory with matter Lagrangian $\mathcal{L}_{M}$ is defined
as
\begin{equation}\label{1}
\mathcal{I}=\int{dx^4\sqrt{-g}\left[\frac{M_p^2}{2}
f(R)+\mathcal{L}_{(M)}\right]}.
\end{equation}
In literature \cite{18}-\cite{22}, people have discussed the
cosmological reconstruction of $f(R)$ theory according to the class
of HDE models. Capozziello et al. \cite{18} developed an effective
numerical scheme for reconstructing $f(R)$ from Hubble parameter of
a given DE model and applied this scheme to the quintessence DE
model and chaplygin gas.

Following \cite{18}, Wu and Zhu \cite{19} reconstructed $f(R)$
according to HDE and explored the future evolution for different
values of the parameter $c$. Feng \cite{20} analyzed the effect of
parameter $\alpha$ on reconstructed $f(R)$ corresponding to Ricci
DE. The explicit functions of $f(R)$ in FRW universe can also be
obtained from the reconstruction procedure according to the given DE
model. Setare \cite{21} obtained $f(R)$ functions corresponding to
HDE and NADE by assuming an ansatz for the scale factor. In
\cite{22}, reconstruction has been executed for both ordinary and
entropy corrected models of holographic and NADE. In a recent work
\cite{16}, we have reconstructed $f(R,T)$ models according to
holographic and NADE and found that the said models can represent
the quintessence/phantom regimes of the universe.

Here, we regard the NADE and NHDE as promising models and apply the
numerical scheme for reconstructing $f(R)$ without introducing any
additional DE factor. The future evolution of $f(R)$ is presented
for different values of the essential parameters. We assure the
evolution of $f(R)$ by analyzing the corresponding behavior of
cosmographic parameters in particular DE models. The paper has the
following format: In section \textbf{2}, we reconstruct the $f(R)$
theory according to NADE and discuss the future evolution. Section
\textbf{3} provides the evolution of $f(R)$ corresponding to NHDE.
In section \textbf{4}, we summarize our findings.

\section{Reconstruction from NADE}

We consider the NADE density of the form \cite{5}
\begin{equation}\label{2}
\rho_{\vartheta}=\frac{3n^2M^{2}_p}{\eta^2},
\end{equation}
where the factor $3n^2$ is inserted to parameterize some
uncertainties namely, the specific forms of cosmic quantum fields,
the role of spacetime curvature \emph{etc.} and $\eta$ is the
conformal time in FRW background
\begin{equation}\label{3}
\eta=\int{\frac{dt}{a(t)}} =\int{\frac{da}{Ha^{2}}}.
\end{equation}
For the flat FRW geometry comprising of matter component and NADE,
the Friedmann equation is given by
\begin{equation}\label{4}
3H^2=\rho_M+\rho_{\vartheta},
\end{equation}
where $\rho_M=\rho_{M0}e^{-3x}$ from the energy conservation
equation of matter. By defining the fractional matter and DE
densities $\Omega_{\vartheta}=\frac{\rho_{\vartheta}}{\rho_{cri}}$,
$\Omega_{M}=\frac{\rho_M}{\rho_{cri}}$ with
$\rho_{cri}=3M^{2}_pH^2$, the Hubble parameter $H(x)$ is obtained as
\begin{equation}\label{5}
H(x)=H_0\left(\frac{\Omega_{M0}e^{-3x}}
{1-\Omega_{\vartheta}}\right)^{1/2}.
\end{equation}
Differentiating $\rho_{\vartheta}$ with respect to $x$ and making
use of DE conservation equation, the equation of state (EoS)
parameter in NADE is obtained as
\begin{equation}\label{6}
\omega_{\vartheta}=-1+\frac{2\sqrt{\Omega_{\vartheta}}e^{-x}}{3n}.
\end{equation}
Using Eqs.(\ref{2}) and (\ref{3}) with the relation
$\Omega_{\vartheta}=\frac{n^2}{H^2{\eta}^2}$, we obtain
\begin{equation}\label{7}
\Omega_{\vartheta}'=\Omega_{\vartheta}(1-\Omega_{\vartheta})
\left(3-\frac{2}{n}\sqrt{\Omega_{\vartheta}}e^{-x}\right),
\end{equation}
where prime denotes derivative with respect to $x=\ln a$. The
initial condition on $\Omega_{\vartheta}$ can be set from
Eq.(\ref{4}) as
\begin{equation}\label{8}
\Omega_{\vartheta0}+\Omega_{M0}=1.
\end{equation}
One can determine $\Omega_{\vartheta}$ using Eqs.(\ref{7}) and
(\ref{8}) and hence the evolution of the universe in NADE can be
executed.

The field equations of $f(R)$ theory can be found by varying action
(1) with respect to the metric
\begin{equation}\label{9}
G_{\alpha\beta}=M^{-2}_pT^{(M)}_{\alpha\beta}+T^{(curv)}_{\alpha\beta},
\end{equation}
where
\begin{equation}\label{10}
T_{\alpha\beta}^{(curv)}=\frac{1}{f_R}\left[\frac{1}{2}g_{\alpha\beta}(f-Rf_R)
+f_R^{;{\mu\nu}}(g_{\alpha\mu}g_{\beta\nu}-g_{\alpha\beta}g_{\mu\nu})\right],
\end{equation}
originates from the curvature contribution to the effective
energy-momentum tensor, the subscript $R$ denotes derivative with
respect to the scalar curvature and
$T_{{\mu}{\nu}}^{(M)}=\hat{T}_{{\mu}{\nu}}^{(M)}/f(R)$,
$\hat{T}_{{\mu}{\nu}}^{(M)}$ is the standard matter energy-momentum
tensor. For the flat FRW geometry, the respective field equations
together with the conservation equation are given by
\begin{eqnarray}\label{11}
&&3M^{2}_pH^2=\rho_T, \quad -M^{2}_p(2\dot{H}+3H^2)=p_T,\\\label{12}
&&\dot{\rho}_T+3H(\rho_T+\omega_T)=0,
\end{eqnarray}
where $\rho_T=\rho_M+\rho_{curv}$ and $p_T=p_M+p_{curv}$. In this
discussion, we consider the pressureless matter without any
curvature-matter interaction. Equations (\ref{11}) and (\ref{12})
can be combined to single equation
\begin{equation}\label{13}
\dot{H}=\frac{-1}{2f}\left[3H_0^2\Omega_{M0}e^{-3x}
+(\ddot{R}-H\dot{R})f_R+\dot{R}^2f_{RR}\right].
\end{equation}
Employing the relation $d/dt=Hd/dx$, we replace $t$ by $x$ and hence
Eq.(\ref{13}) can be translated into 3rd order differential equation
of $f(x)$ as
\begin{equation}\label{14}
\mathcal{B}_3(x)\frac{d^3f}{dx^3}+\mathcal{B}_2(x)
\frac{d^2f}{dx^2}+\mathcal{B}_1(x)\frac{df}{dx}
=-3H^2_0\Omega_{M0}e^{-3x},
\end{equation}
where $\mathcal{B}_i$ are functions of $H(x)$ and its derivatives
given by (\ref{A1}).

We aim to solve this equation to obtain $f[R(x)]$ using the Hubble
parameter $H(x)$. For this purpose, we set the boundary conditions
of the form \cite{18}
\begin{eqnarray}\label{15}
&&\left(\frac{df}{dx}\right)_{x=0}=
\left(\frac{dR}{dx}\right)_{x=0}, \quad
\left(\frac{d^2f}{dx^2}\right)_{x=0}=
\left(\frac{d^2R}{dx^2}\right)_{x=0},\\\label{17}
&&f(x=0)=f(R_0)=6H_0^2(1-\Omega_{M0})+R_0.
\end{eqnarray}
If the function $H(x)$ is known then the coefficients
$\mathcal{B}_i$ and hence the function $f(R)$ can be evaluated
corresponding to given DE model. In case of NADE, we do not have
explicit form of $H(x)$ whereas $H(x)$ and its derivatives can be
represented in terms of $\Omega_{\vartheta}(x)$. Therefore, after
lengthy calculations, the coefficients $\mathcal{B}_i$ are
interpreted in the form of $\Omega_{\vartheta}(x)$ and
$\Omega'_{\vartheta}(x)$. We solve numerically the system of
equations (\ref{7}) and (\ref{14}) together with the conditions
(\ref{8}), (\ref{15}) and (\ref{17}).
\begin{figure}
\centering \epsfig{file=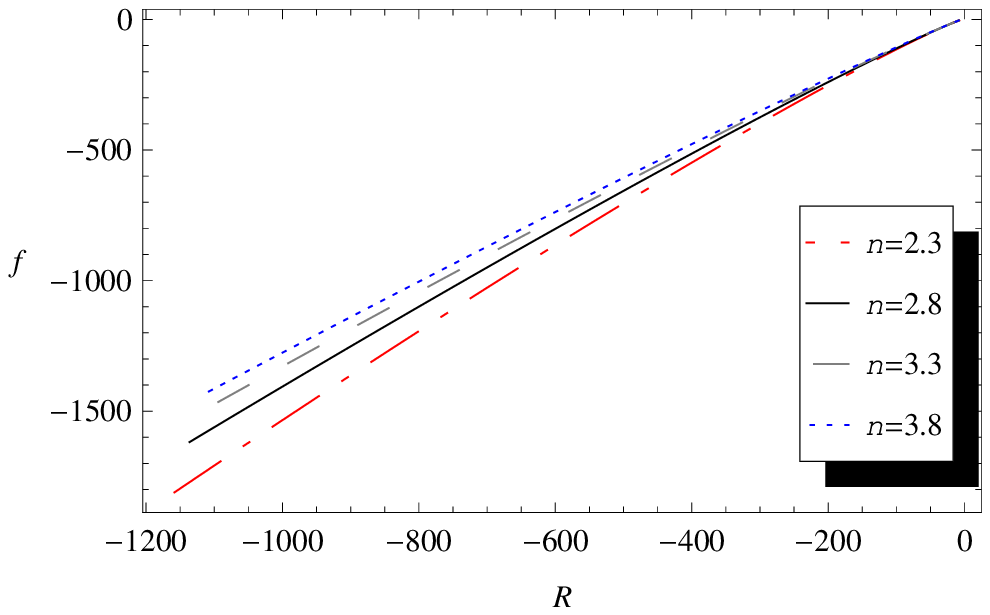} \caption{Reconstructed $f(R)$ for
NADE with $0\leqslant{z}\leqslant10$.}
\end{figure}
\begin{figure}
\centering \epsfig{file=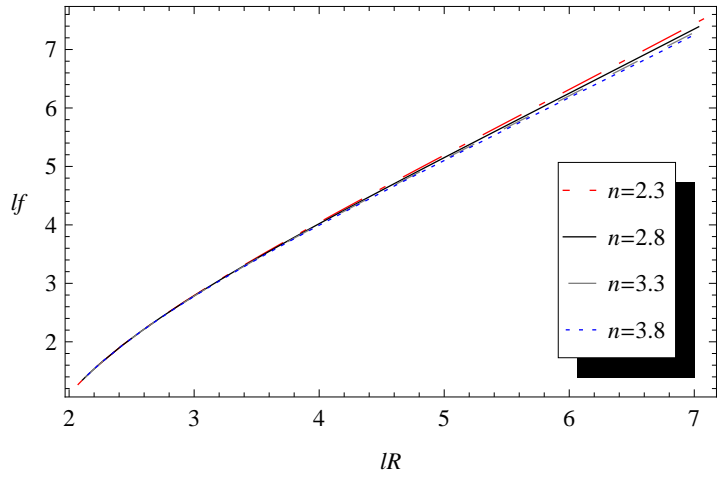} \caption{Reconstructed $f(R)$ for
NADE in $lf-lR$ plane with $0\leqslant{z}\leqslant10$.}
\end{figure}

In \cite{23}, the constraint on parameter $n$ is developed from the
cosmological data for a flat universe consisting of DE and matter
component, the best fit value is found to be
$n=2.76^{+0.111}_{-0.109}$. For the non-flat universe, Zhang et al.
\cite{24} found that the most appropriate measure of $n$ from the
WMAP 7-yr observations is
$n=2.673^{+0.053+0.127+0.199}_{-0.077-0.151-0.222}$. In this study,
we set $n=2.3,~2.8,~3.3,~3.8$ and $\Omega_{M0}=0.27$. For this
choice of parameters, the function $f(R)$ is plotted against $R$ as
shown in Figure \textbf{1}. It is clear that functions appear
distinct if $|R|$ (or $z=e^{-x}-1$) is large, while these functions
seem to coincide for small $|R|$. The evolution of $f$ shown in
Figure \textbf{1} is quite similar to that for HDE \cite{19}. We
also plot these functions on $lf-lR$ plane as shown in Figure
\textbf{2}, where $lf=\ln(-f)$ and $lR=\ln(-R)$. Our results are
consistent with that in \cite{18} and the parameter $n$ is appeared
as fundamental element in identifying the nature of NADE.
\begin{figure}
\centering \epsfig{file=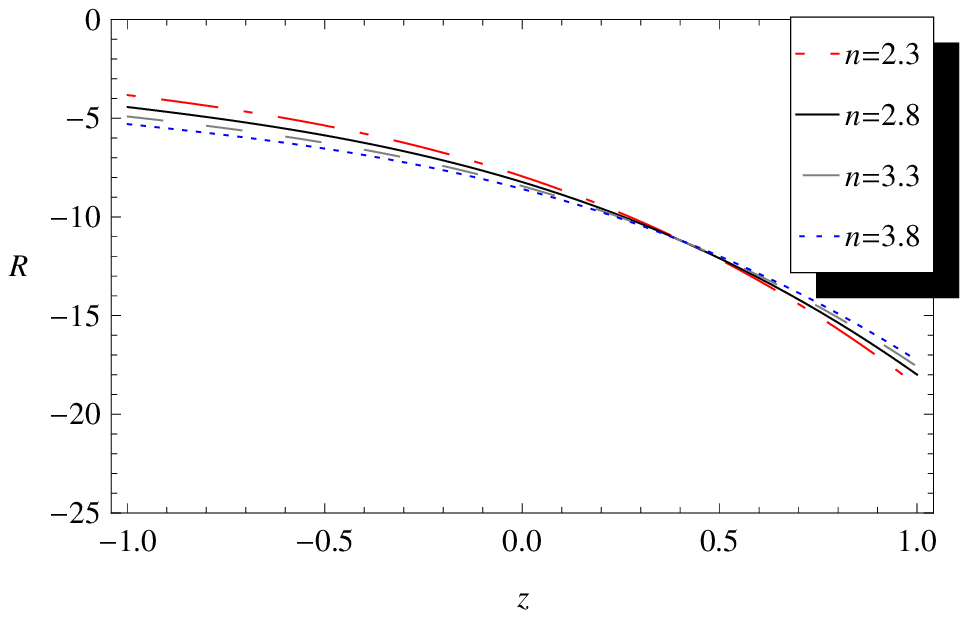} \caption{Future evolution of $R$
in NADE with $-1\lesssim{z}\leqslant1$.}
\end{figure}
\begin{figure}
\centering \epsfig{file=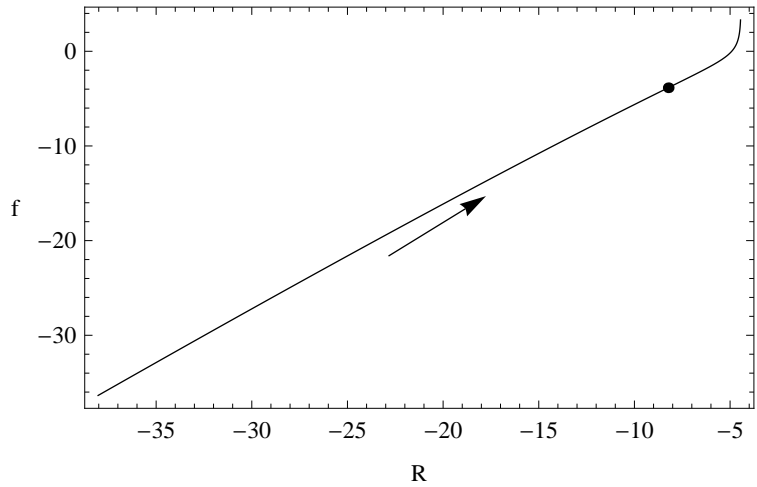} \caption{Future evolution of
$f(R)$ for $n=2.8$ with $-1\lesssim{z}\leqslant2$.}
\end{figure}

To explore the effect of $n$ further, let us see the future
evolution of $|R|$. Figure \textbf{3} shows that future variation of
$|R|$ is alike for different values of $n$ and almost favors the
cosmological constant. In Figures \textbf{4} and \textbf{5}, we
present the future evolution of $f$ for $n=2.3,~2.8,~3.3,~3.8$. For
$n=2.8$, the curve seems to be similar to that for $c=1$ in HDE but
in the late stage of the universe there would be a sudden change.
The evolution of $f$ is similar for other values of $n$ which is
shown in Figure \textbf{5}. These plots indicate distinct
characteristics for the NADE.
\begin{figure}
\centering \epsfig{file=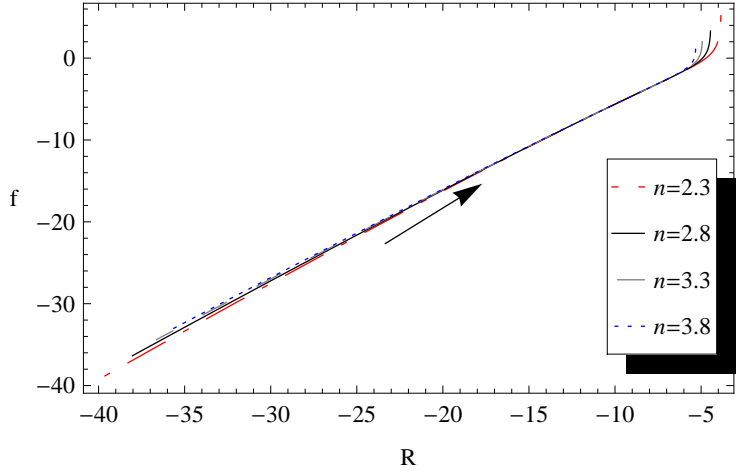} \caption{Future evolution of
$f(R)$ for different values of $n$ with $-1\lesssim{z}\leqslant2$.}
\end{figure}
\begin{figure}
\centering \epsfig{file=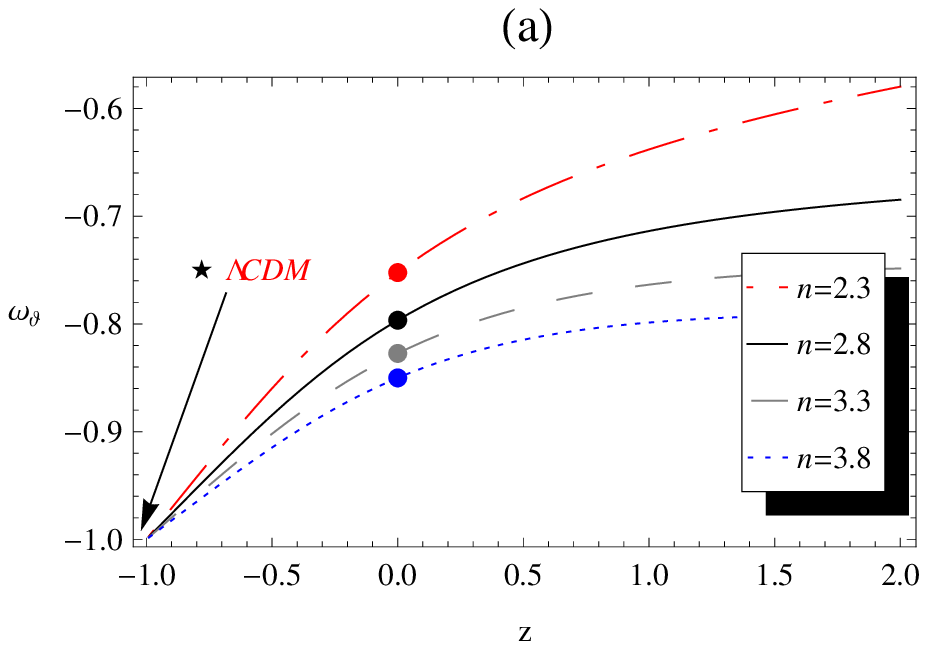,width=.495\linewidth,
height=2.2in} \epsfig{file=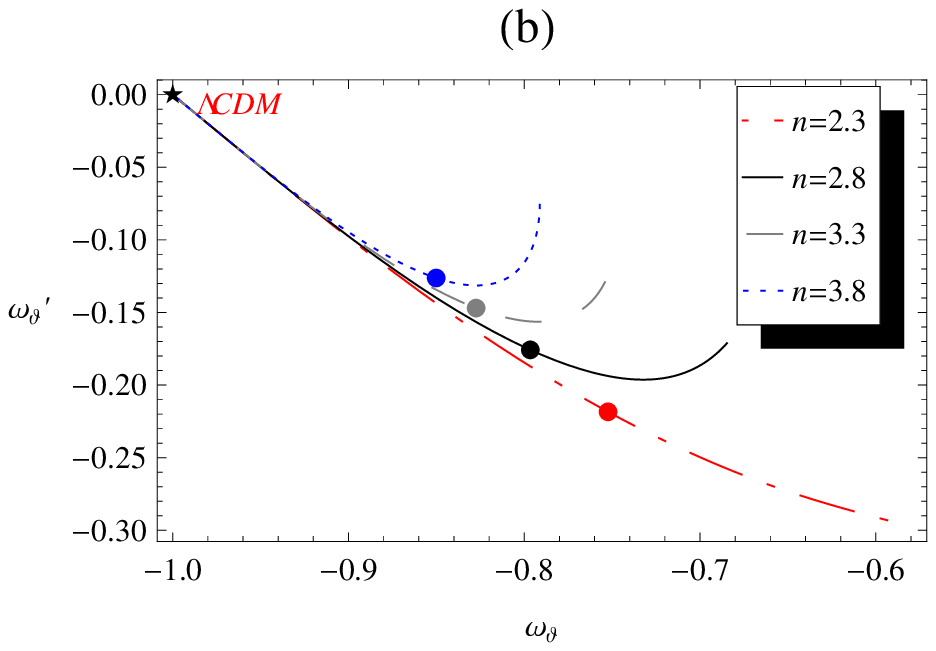, width=.495\linewidth,
height=2.2in} \caption{ Evolution trajectories in NADE for (a)
$\omega_{\vartheta}$ vs $z$ and (b) $\omega_{\vartheta}$ and
$\omega_{\vartheta}'$ in $\omega_{\vartheta}-\omega_{\vartheta}'$
plane with $-1\lesssim{z}\leqslant2$.}
\end{figure}

We extend our discussion and explore the evolution of NADE for the
cosmographic parameters such as EoS parameter, deceleration
parameter, statefinder diagnostic and
$\omega_{\vartheta}-\omega'_{\vartheta}$ analysis. It is clear from
Eq.(\ref{6}) that NADE does not permit the crossing of phantom
divide line $\omega_{\vartheta}=-1$, also if $z\rightarrow-1$ and
$\Omega_{\vartheta}\rightarrow1$ then
$\omega_{\vartheta}\rightarrow-1$ in future evolution of NADE.
Figure \textbf{6(a)} shows that for our choice of parameter $n$, the
EoS parameter of NADE favors the quintessence era and in the late
time, it mimics the cosmological constant regime. We also show the
evolution of $\omega_{\vartheta}'$ in
$\omega_{\vartheta}-\omega_{\vartheta}'$ plane for different values
of $n$ in NADE. Figure \textbf{6(b)} depicts that the
$\omega_{\vartheta}-\omega_{\vartheta}'$ plane represents the
$\Lambda$CDM model $(\omega_{\vartheta}=-1,~\omega_{\vartheta}'=0$)
when $z\rightarrow-1$ (or $x\rightarrow{\infty}$). The present
values of $\omega_{\vartheta}$ and $\omega_{\vartheta}'$ are denoted
by dots on each curve. For $n=2.3,~2.8,~3.3$ and $3.8$, the present
values of ($\omega_{\vartheta},~\omega_{\vartheta}'$) are given by
$(-0.752,-.218),~(-0.796,-.175),~(-0.827,-.147)$ and
$(-0.850,-.126)$, respectively.

Sahni et al. \cite{25} defined the statefinder diagnostic parameters
$\{r,s\}$ of the form
\begin{eqnarray}\label{18}
r=\frac{\dddot{a}}{aH^3},\quad s=\frac{(r-1)}{3(q-1/2)}.
\end{eqnarray}
Introducing the EoS parameter and dimensionless density of DE,
Eq.(\ref{18}) is transformed as
\begin{eqnarray}\label{19}
r&=&1-\frac{3}{2}\Omega_{\vartheta}\left[\omega_{\vartheta}'
-3\omega_{\vartheta}(1+\omega_{\vartheta})\right], \\\label{20}
s&=&\frac{-1}{3\omega_{\vartheta}}\left[\omega_{\vartheta}'
-3\omega_{\vartheta}(1+\omega_{\vartheta})\right].
\end{eqnarray}
The deceleration parameter $q$ in terms of $\omega_{\vartheta}$ and
$\Omega_{\vartheta}$ is given by
\begin{eqnarray}\label{21}
q=\frac{1}{2}(1+3\omega_{\vartheta}\Omega_{\vartheta}).
\end{eqnarray}
The variation of deceleration parameter $q$ with $z$ for the NADE
without interaction is shown in Figure \textbf{7(a)}. The transition
of the universe from decelerating epoch to the accelerated era can
be seen and it will end up with $q=-1$ representing the de Sitter
model. The sign flip of $q$ depends on the selection of $n$, the era
of cosmic acceleration starts earlier for small values of $n$ as
compared to larger values.

The plots of statefinder parameters in the $s-r$ plane for
$n=2.3,~2.8,~3.3$ and $3.8$ are shown in Figure \textbf{7(b)}. The
dots in the diagram correspond to present day values of statefinder
parameters $(r_0,~s_0)$ which are denoted as $(0.150, 0.627)$(red),
$(0.129, 0.660)$(black), $(0.113, 0.691)$(gray) and $(0.100, 0.720)$
(blue). The evolution trajectories of the statefinder diagnostic are
represented for the future evolution and these will end up to star
symbol $\{r=1,s=0\}$, the \emph{$\Lambda$CDM} model. We also plot
the statefinder diagnostic in $q-r$ plane for our selection of
parameter $n$ together with the flat \emph{$\Lambda$CDM} model. It
can be seen from Figure \textbf{8} that evolution trajectories for
NADE in $q-r$ plane commence from the fix point $(q=0.5,~r=1)$ which
represents the standard cold dark matter regime (\emph{SCDM}). These
curves end at $(q=-1,~r=1)$, the de Sitter model in future evolution
of the universe. The past and future eras of the universe and
present day values of $(q,~r)$ are represented by stars and dots,
respectively.
\begin{figure}
\centering \epsfig{file=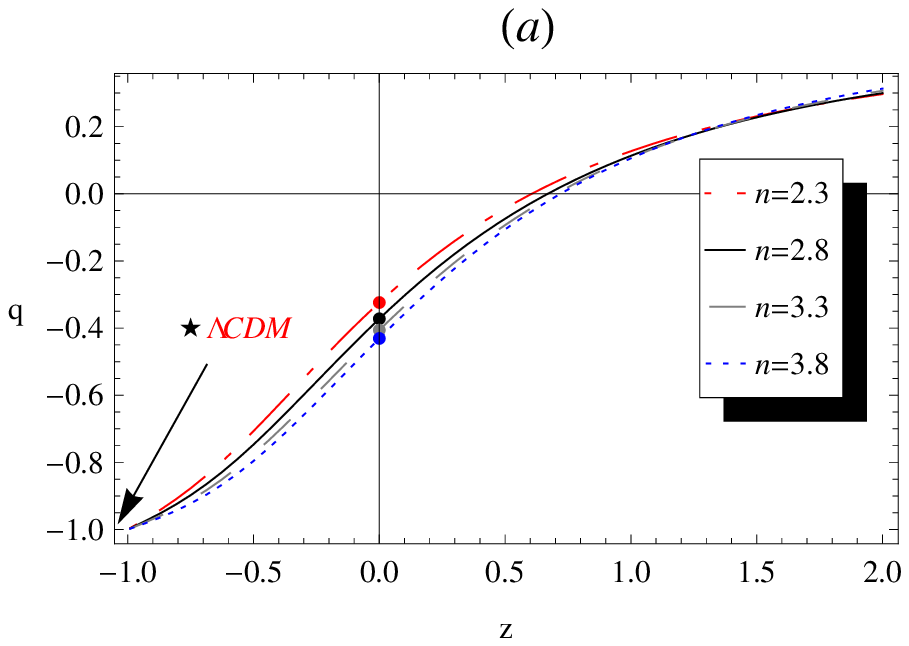,width=.495\linewidth,
height=2.2in} \epsfig{file=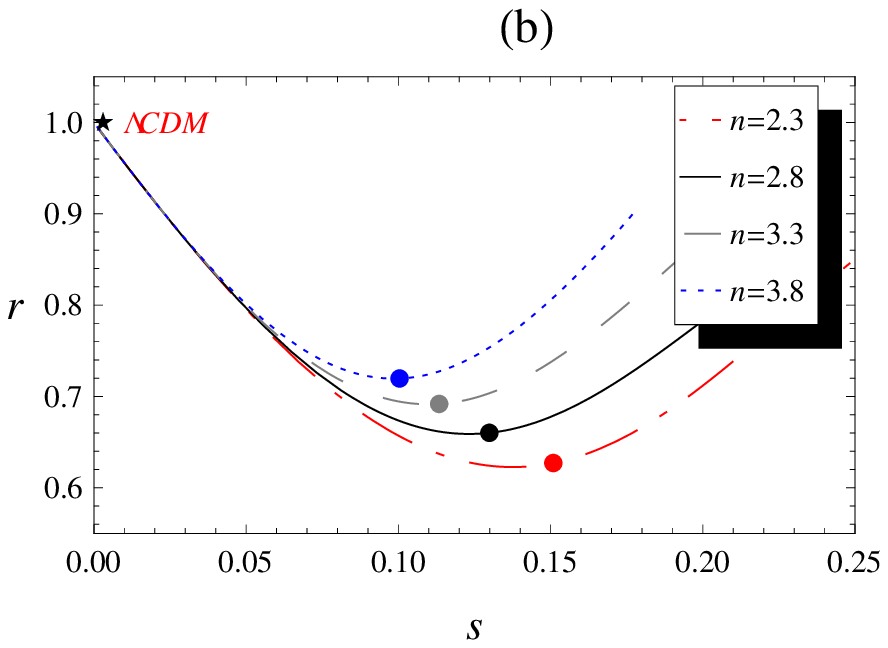,width=.495\linewidth,
height=2.2in} \caption{Evolution trajectories in NADE for (a) $q$ vs
$z$ and (b) the statefinder diagnostic in $s-r$ plane with
$-1\lesssim{z}\leqslant2$.}
\end{figure}
\begin{figure}
\centering \epsfig{file=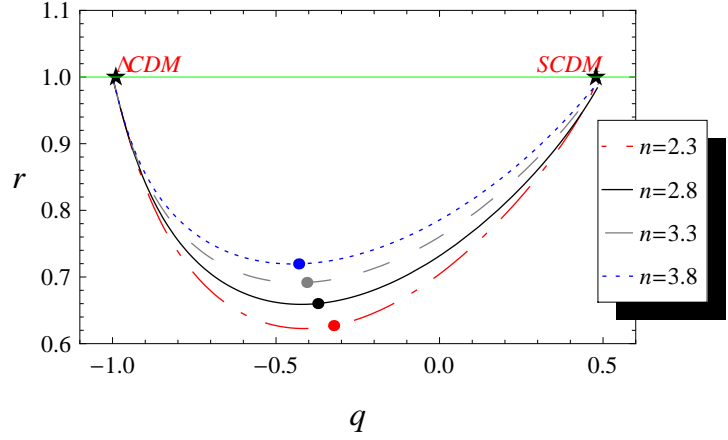} \caption{The statefinder plot for
NADE in $q-r$ plane. The green line represents the $\Lambda$CDM with
$-1\lesssim{z}\leqslant2$.}
\end{figure}

\section{Reconstruction from NHDE}

The energy density of HDE with Granda-Oliveros cutoff is given by
\cite{9}
\begin{equation}\label{22}
\rho_{\vartheta}=3M^{2}_p(\mu{H}^2+\upsilon\dot{H}),
\end{equation}
where $\mu$ and $\upsilon$ are positive constants. Using this value
of $\rho_{\vartheta}$, Eq.(\ref{4}) can be written in the form
\begin{equation}\label{23}
E^2(x)=\frac{2}{2(1-\mu)+3\upsilon}\Omega_{M0}e^{-3x}
+g_0e^{\frac{2x(1-\mu)}{\upsilon}},
\end{equation}
where $E(x)=H(x)/H_0$, $H_0$ is the present day value of Hubble
parameter and $g_0$ is the constant of integration which can be
obtained using the condition $E(x=0)=1$ as
\begin{equation}\label{24}
g_0=1-\frac{2}{2(1-\mu)+3\upsilon}\Omega_{M0}.
\end{equation}
Following \cite{9}, the NHDE density is expressed as
\begin{equation}\label{25}
\rho_{\vartheta}=3H_0^2\left[\frac{2\mu-3\upsilon}{2(1-\mu)+3\upsilon}\Omega_{M0}e^{-3x}
+g_0e^{\frac{2x(1-\mu)}{\upsilon}}\right].
\end{equation}
The pressure of NHDE can be obtained using this value of
$\rho_{\vartheta}$ in the conservation equation of DE
\begin{eqnarray}\label{26}
p_{\vartheta}&=&-\rho_{\vartheta}-\frac{1}{3}\frac{d{\rho}_{\vartheta}}{dx}
=-3H_0^2\left(\frac{2(1-\mu)+3\upsilon}{3\upsilon}\right)
g_0e^{\frac{2x(1-\mu)}{\upsilon}}.
\end{eqnarray}
Manipulating Eqs.(\ref{25}) and (\ref{26}), the EoS parameter of
NHDE turns out to be
\begin{eqnarray}\label{27}
\omega_{\vartheta}=-\frac{[2(1-\mu)+3\upsilon]^2g_0e^{\frac{2x(1-\mu)}{\upsilon}}}
{3\upsilon(2\mu-3\upsilon)\Omega_{M0}e^{-3x}
+[2(1-\mu)+3\upsilon]g_0e^{\frac{2x(1-\mu)}{\upsilon}}}.
\end{eqnarray}
Proceeding in a similar fashion as in the case of NADE, we obtain
\begin{equation}\label{28}
\mathcal{D}_3(x)\frac{d^3f}{dx^3}+\mathcal{D}_2(x)
\frac{d^2f}{dx^2}+\mathcal{D}_1(x)\frac{df}{dx}
=-\Omega_{M0}e^{-3x},
\end{equation}
where $\mathcal{D}_i$ are functions of $E(x)$ and its derivatives,
see Appendix (\ref{A2}). For the HDE with Granda-Oliveros cutoff,
the expression for $H(x)$ is directly useable in numerical
computations. If we substitute Eq.(\ref{23}) and constraint
(\ref{24}) in differential equation (\ref{28}), then the resulting
equation can be solved numerically under the boundary conditions
(\ref{15}) and (\ref{17}).

In \cite{9}, the best fit values of parameters $\mu$ and $\upsilon$
are suggested as $\mu\approx0.93$ and $\upsilon\approx0.5$ to keep
NHDE consistent with the theory of big-bang nucleosynthesis. Wang
and Xu \cite{26} developed the best fit values of parameters
$(\mu,\upsilon)$ in both flat and non-flat NHDE models from the
current observational data. They found the best fit parameters for
the flat model as $\mu=0.8502^{+0.0984+0.1299}_{-0.0875-0.1064}$ and
$\upsilon=0.4817^{+0.0842+0.1176}_{-0.0773-0.0955}$. In this study,
we select the parameters $(\mu=0.85, \upsilon=0.48)$, $(\mu=0.93,
\upsilon=0.56)$ and $(\mu=1, \upsilon=0.63)$. The plot of the
function $f(R)$ versus $R$ for NHDE is shown in Figure
\textbf{9(a)}. It shows that reconstructed $f(R)$ for NHDE is the
same for different values of parameters $(\mu,\upsilon)$. We also
plot these results on $\emph{lf}-\emph{lR}$ plane as shown in Figure
\textbf{9(b)}. To be more definite about the behavior of $f$, we
draw plot on $\emph{lf}-x$ plane presented in Figure \textbf{10}.
This shows that different values of parameters do not affect the
shapes of curves unless $x>-0.3$ and aftermath these curves would
depict different picture.
\begin{figure}
\centering \epsfig{file=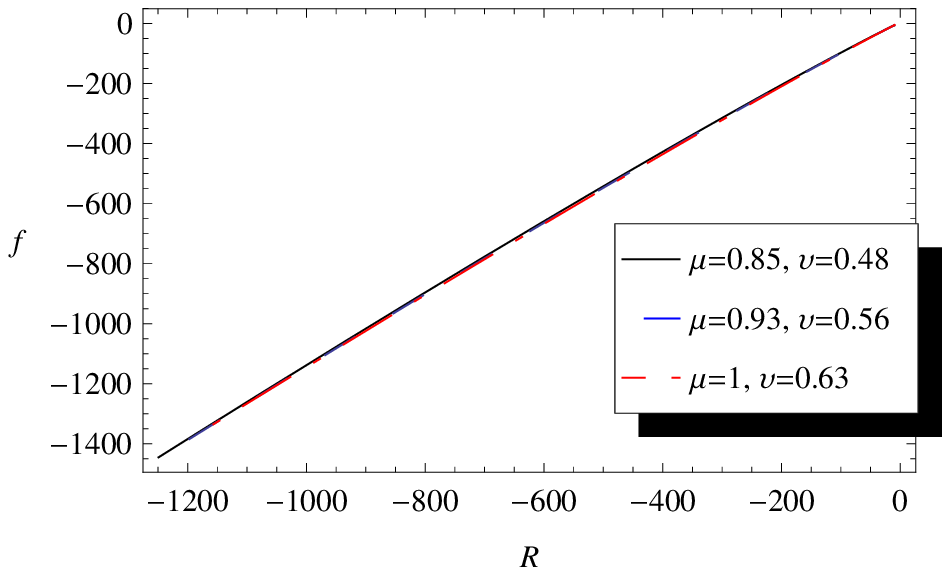,width=.495\linewidth,
height=2.2in}\epsfig{file=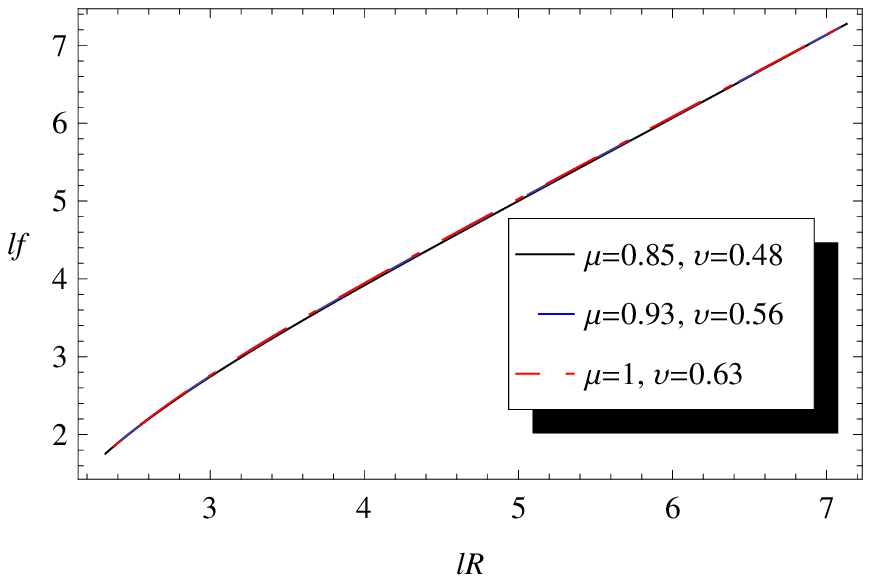,width=.495\linewidth,
height=2.2in} \caption{Reconstructed $f(R)$ in (a) $f-R$ plane and
(b) $lf-lR$ plane for NHDE with $0\leqslant{z}\leqslant10$.}
\end{figure}
\begin{figure}
\centering \epsfig{file=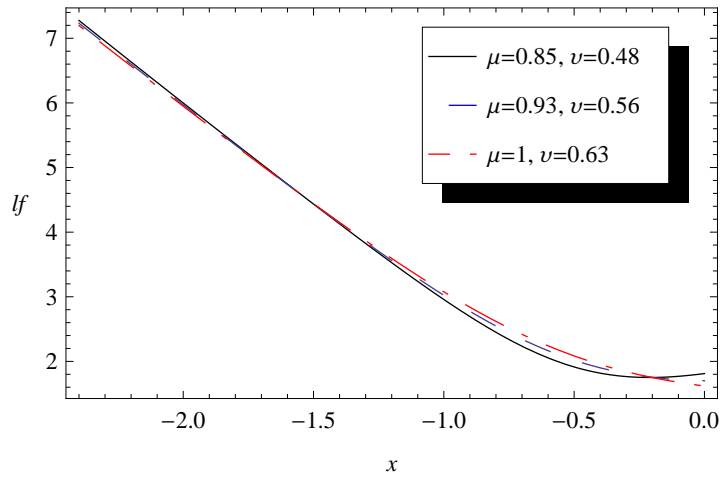} \caption{Reconstructed $f(R)$ for
NHDE in $lf-x$ plane with $0\leqslant{z}\leqslant10$, where
$x=\ln(1+z)^{-1}$.}
\end{figure}
\begin{figure}
\centering \epsfig{file=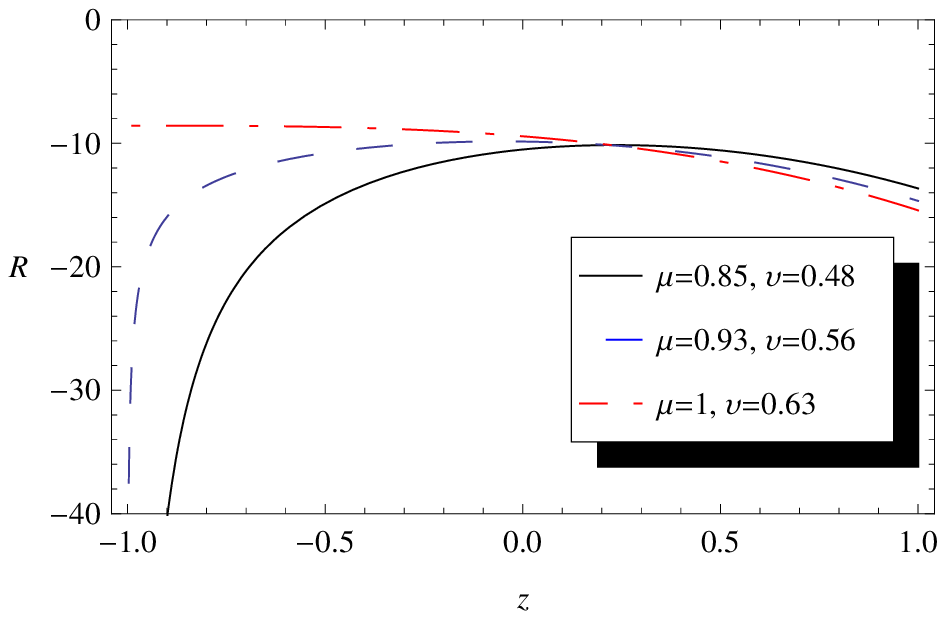} \caption{Future evolution of $R$
in NHDE with $-1\leqslant{z}\leqslant1$.}
\end{figure}

In order to explore the distinctive effect of parameters
$(\mu,\upsilon)$, we get insight of future evolution. First we
investigate the future evolution of $R$ versus red shift which is
shown in Figure \textbf{11}. For $(\mu, \upsilon)<(1, 0.63)$, $|R|$
would take infinitely large values in future which indicate the
phantom era with $\omega_{\vartheta}<-1$ dominating over the matter
part, leading to the big rip singularity. When $(\mu, \upsilon)=(1,
0.63)$, there is a slight variation in $|R|$ and diagram assures the
DE model with $\omega_{\vartheta}=-1$, the cosmological constant.
\begin{figure}
\centering \epsfig{file=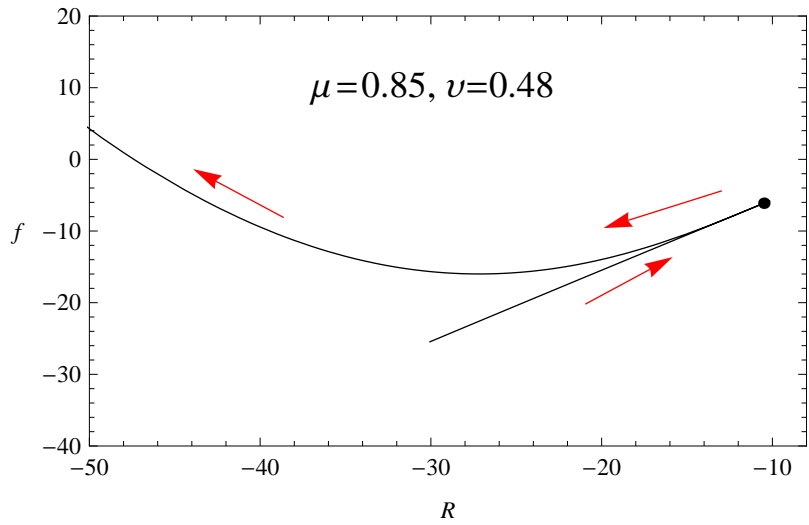} \caption{Future evolution of
$f(R)$ for $(\mu,\upsilon)=(0.85,0.48)$ with
$-1\lesssim{z}\leqslant2$. The dot denotes the present day and is
the point of reversion for DE dominated model.}
\end{figure}
The difference in the selected parameters $(\mu, \upsilon)$ can be
seen more effectively in the future evolution of $f$ reconstructed
according to the NHDE as shown in Figures
\textbf{(12)}-\textbf{(14)}. For $\mu=0.85$ and $\upsilon=0.48$ in
Figure \textbf{12}, the curve shows that initially $|R|$ decreases
before reaching the present epoch $(x=0)$ which changes its
direction and it would increase leading to the phantom DE. In this
scenario, $|R|$ keeps growing whereas $f$ initially decreases and
then attains positive value approaching to $+\infty$. In fact, in
phantom DE models the point of reversion is a common characteristic
because the DE components succeed in their competition with matter
contents of the universe. For $\mu=0.93$ and $\upsilon=0.56$, we
have almost identical picture as in Figure \textbf{12} but here
growing rate in $f$ is comparatively large. For $\mu=1$ and
$\upsilon=0.63$, we have a linear dependence of $f$ on $R$ leading
to constant which is in accordance to the de Sitter model, where
$f(R)=R+\Lambda$.
\begin{figure}
\centering \epsfig{file=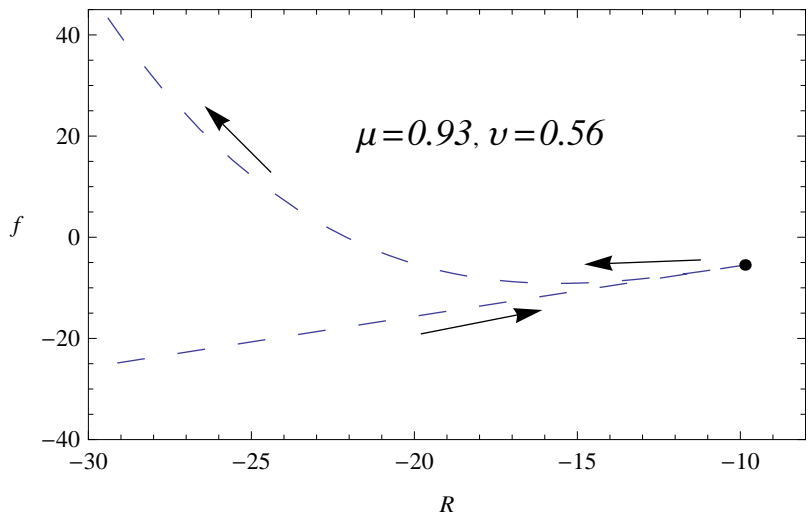} \caption{Future evolution of
$f(R)$ for $(\mu,\upsilon)=(0.93,0.56)$ with
$-1\lesssim{z}\leqslant2$. The dot denotes the present day and is
the point of reversion for DE dominated model.}
\end{figure}
\begin{figure}
\centering \epsfig{file=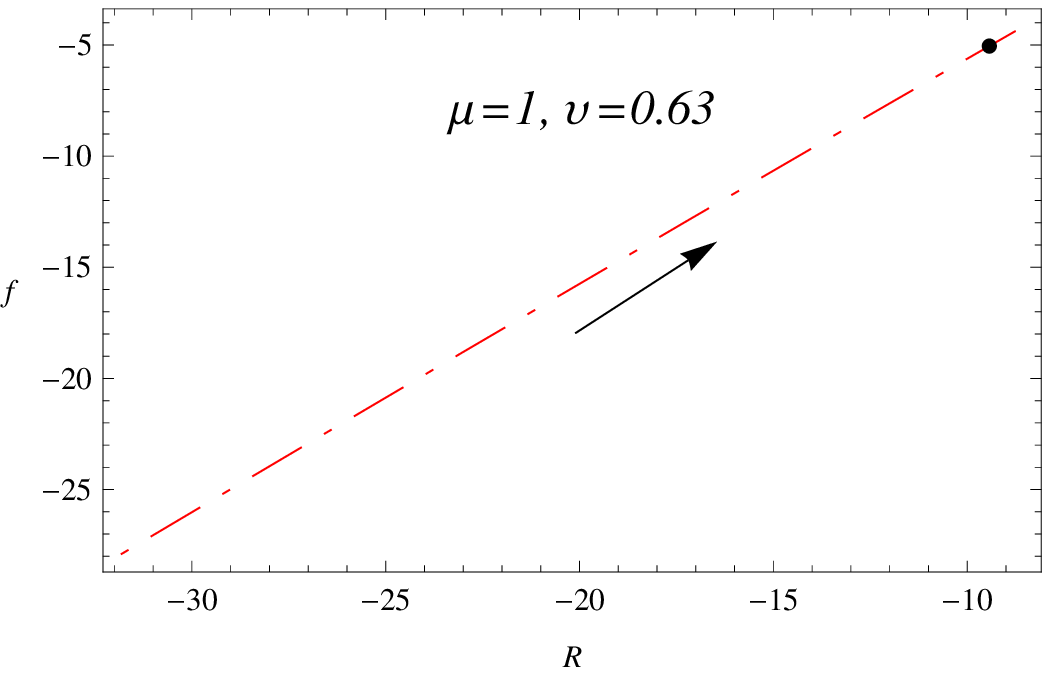} \caption{Future evolution of
$f(R)$ for $(\mu,\upsilon)=(1,0.63)$ with $-1\lesssim{z}\leqslant2$.
The dot denotes the present day and is the point of reversion for DE
dominated model.}
\end{figure}
\begin{figure}
\centering \epsfig{file=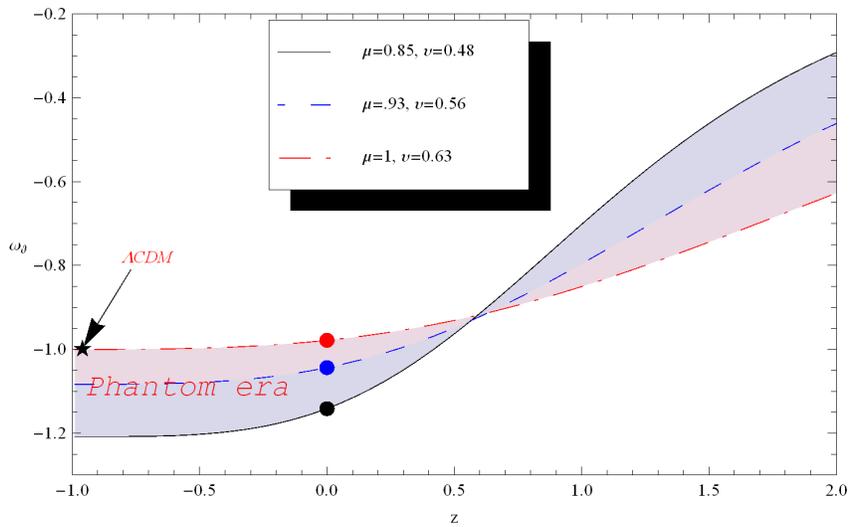, width=.85\linewidth,height=2.8in}
\caption{Evolution trajectories of EoS parameter in NHDE for
different values of parameters $(\mu,\upsilon)$ with
$-1\lesssim{z}\leqslant2$. The dot denotes the present day and star
represents the \emph{$\Lambda$CDM} model.}
\end{figure}

Now we discuss the evolution of the NHDE for the selected parameters
$(\mu, \upsilon)$ and interpret the behavior of EoS parameter,
deceleration parameter and statefinder diagnostic. The plot of EoS
parameter for future evolution in NHDE is shown in Figure
\textbf{15}. It shows that the NHDE represents the de Sitter phase
of the universe for $(\mu, \upsilon)=(1,0.63)$. For $(\mu,
\upsilon)<(1,0.63)$, the EoS parameter intersects the phantom divide
line ($\omega_{\vartheta}=-1$) and behaves as quintom model of DE
\cite{26a}. In this perspective, $\omega_{\vartheta}$ ends up with
phantom era which may lead to cosmic doomsday when all the
astronomical objects will be ripped apart. It is evident that domain
of $\omega_{\vartheta}$ in NHDE is consistent with the observational
data of WMAP5 which establishes range of
$-1.11<\omega_{\vartheta}<-0.86$ \cite{27}.

The evolution of $q$ is represented in Figure \textbf{16} which
confirms the behavior of $\omega_\vartheta$. The curve for $(\mu,
\upsilon)=(1,0.63)$ assures the \emph{$\Lambda$CDM} model with
$q=-1$. The transition from deceleration to accelerated epoch can be
seen from this plot and the values of redshift at the transition
point are consistent with the observational results \cite{28}. The
plot of statefinder diagnostic in NHDE for different values of
parameters $(\mu, \upsilon)$ is shown in Figure \textbf{17}. The dot
represents the fix point ${r=1, s=0}$ (\emph{i.e.}, the de Sitter
phase) and all the curves pass through this point.
\begin{figure}
\centering \epsfig{file=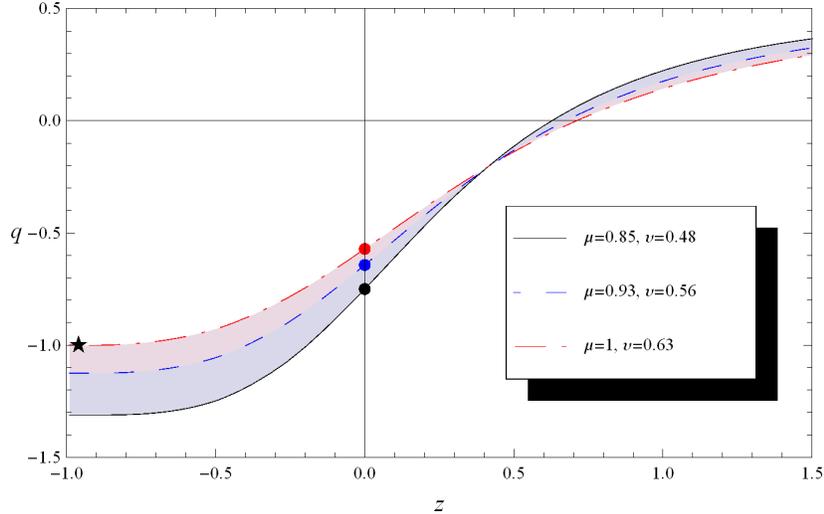, width=0.85\linewidth,
height=2.8in} \caption{Evolution trajectories of $q$ in NHDE for
different values of parameters $(\mu,\upsilon)$ with
$-1\lesssim{z}\leqslant2$. The dot denotes the present day and star
represents the \emph{$\Lambda$CDM} model.}
\end{figure}
\begin{figure}
\centering \epsfig{file=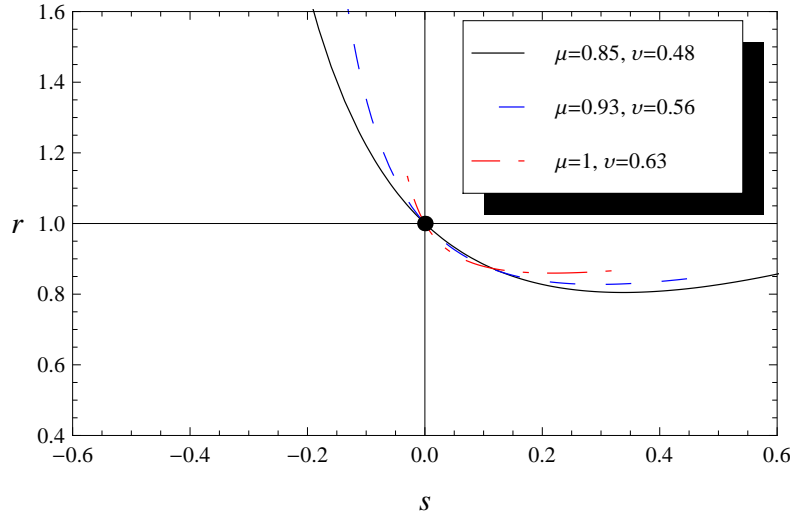, width=0.85\linewidth,
height=2.8in} \caption{Evolution trajectories of statefinder
diagnostic in NHDE for different values of parameters
$(\mu,\upsilon)$ with $-1\lesssim{z}\leqslant2$.}
\end{figure}

\section{Conclusions}

The $f(R)$ theory stands as one of the prosperous contexts to
describe the cosmic evolution and the present day observational
consequences. This theory appears to be a potential candidate in
explaining the late time accelerated expansion. A profound model
which can explain the cosmic evolution in a definite way is still
under consideration. The cosmological reconstruction of $f(R)$
gravity has been explored in \cite{18}-\cite{22} and the issue of
which approach should be used is still alive. In refs.\cite{21,22},
$f(R)$ function corresponding to a class of HDE models has been
constructed by assuming some ansatz for the scale factor in FRW
background. A more effective scheme to reconstruct $f(R)$ theory
from the given evolution history $H(z)$ is developed by Capozziello
et al. \cite{18}. In this scheme, the significant thing is that we
can develop the correspondence of $f(R)$ theory to the given DE
model by using the expression of respective $H(z)$. As a result, one
can find the $f(R)$ theory which explains the same dynamics
(\emph{i.e.}, the cosmic evolution) as predicted by the given DE
model. Now it is of interest to consider the modified HDE models and
address the intrinsic degeneracy among the $f(R)$ theory and DE
models. We find that predictions of both candidates ($f(R)$ theory
and DE models) reconcile as they represent distinct features of the
same picture.

In this work, we have reconstructed the function $f(R)$ according to
NADE and NHDE in flat FRW geometry. The numerical reconstruction
scheme is applied to obtain the evolution trajectories of $f$ in
different scenarios. In this reconstruction procedure, the Hubble
parameter plays a significant role as we transform all quantities in
terms of $H$ and $\dot{H}$. We summarize our results as follows:
{\begin{itemize}
\item {For NADE, $H$ is given in terms of
$\Omega_{\vartheta}$, so we solve the system of evolution equations
for both $\Omega_{\vartheta}$ and $f$. The results are shown in
Figures \textbf{1} and \textbf{2} which are consistent with the
constructed functions in literature \cite{18}-\cite{20}. In
comparison with HDE, the future variation of $|R|$ and $f$ show
identical behavior for different values of $n$. We can say that the
behavior of $f$ suggest the de Sitter phase in late time evolution
of the universe. The cosmological parameters have been explored in
NADE for $n=2.3,2.8,3.3,3.8$ to make sure the evolution of $f$.
Figures \textbf{5}-\textbf{8}} evidently show that NADE favors the
quintessence regime and in future evolution, it may end up with the
de Sitter phase. Thus our results for reconstructed $f(R)$ are
consistent with the independent evolution of NADE. We would like to
emphasize that we have taken significantly different values of $n$
but all of these contribute similar results.
\item{In case of HDE with Granda-Oliveros cutoff,
the Hubble parameter in the form $E(x)=H(x)/H_0$ is directly used in
numerical calculations. We have shown the function $f$ in $f-R$ and
$lf-lR$ planes in Figure \textbf{9}}. These curves seem to be
identical for different values of parameters $(\mu, \upsilon)$ and
slight difference is found for $lf-x$ plane which is shown in Figure
\textbf{10}. Further, we probe the future evolution of $|R|$ and
obtain distinct variations accordingly as $(\mu,
\upsilon)\leqslant(1, 0.63)$. The future evolution of $f$ in Figures
\textbf{12}-\textbf{14} evidently show the role of parameters $(\mu,
\upsilon)$. These plot represent distinct features of $f$ which have
been later confirmed by the evolution trajectories of
$\omega_{\vartheta}$ and $q$ in Figures \textbf{15} and \textbf{16}.
For $(\mu, \upsilon)=(0.85, 0.48)~\&~(0.93, 0.56)$, it can be seen
that $f$ depicts the phantom DE era and in such case
$\omega_{\vartheta}<-1$ and $q<-1$. For $(\mu, \upsilon)=(1, 0.63)$,
we have $f$ representing the de Sitter phase with
$\omega_{\vartheta}=-1$ and $q=-1$.  Thus, our results for the
function $f$ corresponding to NHDE coincide with that of
cosmographic parameters.
\end{itemize}}
It is to be noted that NADE and NHDE models are developed in the
context of general relativity rather than any modified theory such
as $f(R)$ gravity. We have reconstructed $f(R)$ by considering the
curvature part as an effective description of these DE models. We
also emphasize that this work is more comprehensive when comparing
with previous ones as it involves the analysis of cosmological
parameters to ensure the evolution of reconstructed function $f(R)$.
\vspace{0.25cm}

{\bf Acknowledgment}

\vspace{0.25cm}

We would like to thank the Higher Education Commission, Islamabad,
Pakistan for its financial support through the {\it Indigenous Ph.D.
5000 Fellowship Program Batch-VII}.

\renewcommand{\theequation}{A.\arabic{equation}}
\setcounter{equation}{0}
\section*{Appendix A}

\begin{eqnarray}\nonumber
\mathcal{B}1&=&2H^2\left(\frac{d^2R}{dx^2}\right)\left(\frac{d^2R}
{dx^2}\right)^{-3}-\left[H^2\frac{d^3R}{dx^3}+
\left(\frac{1}{2}\frac{dH^2}{dx}-H^2\right)\frac{d^2R}{dx^2}\right]
\left(\frac{dR}{dx}\right)^{-2}\\\nonumber&+&\frac{dH^2}{dx}
\left(\frac{dR}{dx}\right)^{-1}, \\\nonumber
\mathcal{B}2&=&-2H^2\left(\frac{d^2R}{dx^2}\right)
\left(\frac{dR}{dx}\right)^{-2}+\left(\frac{1}{2}\frac{dH^2}
{dx}-H^2\right)\left(\frac{dR}{dx}\right)^{-1},
\\\label{A1}
\mathcal{B}3&=&H^2\left(\frac{dR}{dx}\right)^{-1},
\end{eqnarray}
\begin{eqnarray}\nonumber
\mathcal{D}1&=&2E^2\left(\frac{d^2R}{dx^2}\right)\left(\frac{d^2R}
{dx^2}\right)^{-3}-\left[E^2\frac{d^3R}{dx^3}+
\left(\frac{1}{2}\frac{dE^2}{dx}-E^2\right)\frac{d^2R}{dx^2}\right]
\left(\frac{dR}{dx}\right)^{-2}\\\nonumber&+&\frac{dE^2}{dx}
\left(\frac{dR}{dx}\right)^{-1}, \\\nonumber
\mathcal{D}2&=&-2E^2\left(\frac{d^2R}{dx^2}\right)
\left(\frac{dR}{dx}\right)^{-2}+\left(\frac{1}{2}\frac{dE^2}
{dx}-E^2\right)\left(\frac{dR}{dx}\right)^{-1},
\\\label{A2}
\mathcal{D}3&=&E^2\left(\frac{dR}{dx}\right)^{-1}.
\end{eqnarray}


\vspace{.25cm}

\begin{thebibliography}{36}


\bibitem{1}Perlmutter, S. et al.: Astrophys. J.
\textbf{517}(1999)565; Spergel, D.N. et al.: Astrophys. J. Suppl.
\textbf{148}(2003)175; Tegmark, M. et al.: Phys. Rev. D
\textbf{69}(2004)103501; Riess, A.G. et al.: Astrophys. J.
\textbf{659}(2007)98; Fedeli, C., Moscardini, L. and Bartelmann, M.:
Astron. Astrophys. \textbf{500}(2009)667.

\bibitem{2}Cohen, A.G., Kaplan, D.B. and Nelson, A.E.: Phys. Rev.
Lett. \textbf{82}(1999)4971.

\bibitem{3}Li, M.: Phys. Lett. B \textbf{603}(2004)1.

\bibitem{4}Hsu, S.D.H.: Phys. Lett. B \textbf{594}(2004)13.

\bibitem{5}Wei, H. and Cai, R.-G.: Phys. Lett. B
\textbf{660}(2008)113.

\bibitem{6}Wu, J.-P., Ma, D.-Z. and Ling, Y.: Phys. Lett. B
\textbf{663}(2008)152.

\bibitem{7}Liu, X.-L. and Zhang, X.: Commun. Theor. Phys.
\textbf{52}(2009)761.

\bibitem{8}Jamil, M. and Saridakis, E.N.: JCAP \textbf{07}(2011)028.

\bibitem{9}Granda, L.N. and Oliveros, A.: Phys. Lett. B
\textbf{669}(2008)275; ibid. \textbf{671}(2009)199.

\bibitem{10}Sharif, M. and Jawad, A.: Eur. Phys. J. C
\textbf{72}(2012)2097.

\bibitem{11}Sotiriou, T.P and Faraoni, V.: Rev. Mod.
Phys. \textbf{82}(2010)451; De Felice, A. and Tsujikawa, S.: Living
Rev. Rel. \textbf{13}(2010)3; Sharif, M. and Zubair, M.: Astrophys.
Space Sci. \textbf{342}(2012)511; Bamba, K. Capozziello, S. Nojiri,
S. and Odintsov, S.D.: Astrophys. Space Sci. \textbf{345}(2012)155.

\bibitem{12}Harko, T., Lobo, F.S.N., Nojiri, S. and Odintsov, S.D.:
Phys. Rev. D \textbf{84}(2011)024020.

\bibitem{13}Sharif, M. and Zubair, M.: JCAP \textbf{03}(2012)028;
ibid. Erratum: \textbf{05}(2012)E01.

\bibitem{14}Sharif, M. and Zubair, M.: J. Phys. Soc. Jpn. \textbf{81}(2012)114005.

\bibitem{15}Sharif, M. and Zubair, M.: J. Phys. Soc. Jpn. \textbf{82}(2013)014002.

\bibitem{16}Sharif, M. and Zubair, M.: \emph{Cosmology of Holographic and New Agegraphic $f(R,T)$
Models}, J. Phys. Society of Jpn. (to appear, 2013).

\bibitem{16a}Sharif, M. and Zubair, M.: \emph{Thermodynamic Behavior of Particular $f(R,T)$
Gravity Models}, J. Exp. Theor. Phys. (to appear, 2013).

\bibitem{17}Cognola, G., Elizalde, E., Nojiri, S., Odintsov, S.D. and S. Zerbini:
Phys. Rev. D \textbf{73}(2006)084007; Sharif, M. and Abbas, G.: J.
Phys. Soc. Jpn. \textbf{82}(2013)034006.

\bibitem{18}Capozziello, S. Cardone, V.F. and Troisi, A.: Phys. Rev. D
\textbf{71}(2005)043503.

\bibitem{19}Wu, X. and Zhu, Z.-H.: Phys. Lett. B \textbf{660}(2008)293.

\bibitem{20}Feng, C.-J.: Phys. Lett. B \textbf{676}(2009)168.

\bibitem{21}Setare, M.R.: Int. J. Mod. Phys. D
\textbf{17}(2008)2219; Astrophys. Space Sci. \textbf{326}(2010)27.

\bibitem{22}Karami, K. and Khaledian, M.S.: JHEP
\textbf{03}(2011)086.

\bibitem{23}Wei, H. and Cai, R.-G.: Phys. Lett. B
\textbf{663}(2008)1.

\bibitem{24}Zhang, J.-F., Li, Y.-H. and Zhang, X.: Eur. Phys. J. C
\textbf{73}(2013)2280.

\bibitem{25}Sahni, V., Saini, T.D., Starobinsky, A.A. and Alam, U.:
JETP Lett. \textbf{77}(2003)201; Pisma Zh. Eksp. Teor. Fiz.
\textbf{77}(2003)249.

\bibitem{26}Wang, Y. and Xu, L.: Phys. Rev. D \textbf{81}(2010)083523.

\bibitem{26a}Wei, H., Cai, R. G. and Zeng, D. F.: Class. Quantum Grav.
\textbf{22}(2005)3189.

\bibitem{27}Komatsu, E. et al.: Astrophys. J. Suppl.
\textbf{180}(2009)330.

\bibitem{28}Ma, Y.-Z.: Nucl. Phys. B \textbf{804}(2008)262; Daly, R.
A. et al.:  The Astrophysical Journal \textbf{677}(2008)1.


\end{thebibliography}
\end{document}